\documentclass[pdflatex,sn-mathphys-num]{sn-jnl}

\usepackage{graphicx}%
\usepackage{multirow}%
\usepackage{amsmath,amssymb,amsfonts}%
\usepackage{amsthm}%
\usepackage[title]{appendix}%
\usepackage{xcolor}%
\usepackage{textcomp}%
\usepackage{manyfoot}%
\usepackage{booktabs}%
\usepackage{booktabs}%
\usepackage[utf8]{inputenc}
\usepackage[T1]{fontenc}
\usepackage{algorithm}%
\usepackage{algorithmicx}%
\usepackage{algpseudocode}%
\usepackage{listings}%

\theoremstyle{thmstyleone}%
\theoremstyle{thmstyletwo}%

\theoremstyle{thmstylethree}%

\begin{document}

\title[Article Title]{Investigating the correlations of bulk properties of hyperon star with dark matter}


\author[1,2]{\fnm{Rashmita} \sur{Jena}}\email{jrashmitaphy@gmail.com}

\author[3]{\fnm{Padmalaya} \sur{Dash}}\email{padmalayaphy@gmail.com}
\equalcont{These authors contributed equally to this work.}

\author*[2]{\fnm{ S.K.} \sur{Biswal}}\email{subratphy@gmail.com}
\equalcont{These authors contributed equally to this work.}

\affil*[1]{\orgdiv{Department of Physics}, \orgname{Fakir Mohan University}, \orgaddress{ \city{Balasore}, \postcode{756019}, \state{Odisha}, \country{India}}}

\affil[2]{\orgdiv{Department of Physics}, \orgname{K. K. S. Women's College}, \orgaddress{ \city{Balasore}, \postcode{756001}, \state{Odisha}, \country{India}}}

\affil[3]{\orgdiv{Department of Physics}, \orgname{SOA University}, \orgaddress{ \city{Bhubaneswar}, \postcode{751003}, \state{Odisha}, \country{India}}}


\abstract{The strong gravitational pull of the neutron star leads to the accretion of dark matter (DM) inside the core of the neutron star. The accretion of DM affects the bulk properties of the neutron star. Here, we study how the accretion of WIMP (Weakly Interacting Massive Particles) dark matter particles affects the $\Delta-$admixed hyperon star's bulk properties specifically mass, radius, tidal deformability, $f-$mode frequency and moment of inertia. The inclusion of dark matter softens the EOS (equation of state) and reduces the maximum possible mass, canonical radius, canonical tidal deformability, and moment of inertia of canonical star. However, the $f-$mode frequency of the canonical star increases. We find a cubical correlation between the dark matter fermi momenta $k_f^{DM}$ and bulk properties of canonical star.}




\maketitle

\section{Introduction}\label{sec1}
 Extensive evidence supports the existence of dark matter in our universe \cite{Zwicky1937,Bahcall1999,Bergstrom2000,BERTONE2005279,vikhlinin06}. 
 Despite this, the precise nature of dark matter remains a matter of ongoing debate. CMB (Cosmic Microwave Background) analysis from radiation, large-scale structures observations, 
 supernovae, and Baryon Acoustic Oscillation (BAO) indicate that dark matter and dark energy constitute approximately 95\% of the cosmos \cite {Ackerman2008,OKS2021}. 
 The nature of dark matter remains a mystery, but various models or candidates have been proposed, including WIMPs (Weakly Interacting Massive Particles)  \cite{Kouvaris11,Rupin14,Quddus2020}, FIMPs (Feebly Interacting Massive Particle) \cite{Bernal17,Hall10}, sterile neutrinos \cite{Dodelson94,BOYARSKY2019}, axions 
 \cite{Duffy2009,Adams:2022}, Klauza-Klein particles \cite{Cheng2002,SERVANT2003}, WIMPzillas \cite{Kolb:1998,Chung98} and PBH (Primordial Black Hole) \cite{Carr16,Villanueva21}, etc., to analyze their effect on the observable universe.

Detection of dark matter particles can be achieved through
direct or indirect experimental methods. Direct
detection methods, such as XENON \cite{Angle08, Angle08101, APRILE2011, Aprile12}, DAMA \cite{Bernabei09, Bernabei10, Bernabei23}, CDMS \cite{CDMS-II, Ahmed11}, LUX-ZEPLIN 
\cite{Aalbers23, Aalbers108}, CRESST \cite{Abdelhameed19, cresst2024}, SENSEI \cite{Barak20, Gu22}, 
EDELWEISS \cite{ARMENGAUD20131, Armengaud2017} and p-type point-contact germanium detector 
\cite{Aalseth11,Li2022ptype}, etc, are designed to observe the interaction between dark matter candidates and normal matter. These experiments typically use highly sensitive detectors like scintillation detectors, ionization detectors, bolometers, liquid xenon detectors, solid-state detectors, and crystal detectors to identify rare events where dark matter particles collide with atomic nuclei.

Indirect experimental searches focus on detecting the annihilation or decay of dark matter. Instruments such as the Fermi Gamma-ray Space Telescope \cite{BRINGMANN2012}, AMS 
\cite {Salati2017,Heisig20,Calore22}, HESS (high-energy stereoscopic system) \cite{inproceedings,Cembranos2013}, VERITAS (very energetic radiation imaging telescope array) \cite{HOLDER2006}, Ice-Cube Neutrino Observatory \cite{IceCube2017, Iovine2021}, XMM-Newton and Chandra X-ray Observatory \cite{Loewenstein2012} are used in these studies. They aim to capture secondary particles such as gamma rays, neutrinos, and X-rays that result from dark matter interactions. These indirect methods provide complementary insights by observing astrophysical phenomena that might signal the presence of dark matter.

Dark matter is accreted into the densely packed neutron star due to its strong gravitational field. Dark matter interacts with conventional matter such as nucleons, protons, and electrons, resulting in either the release of energy through annihilation or the scattering of the conventional particles via the transfer of energy and momentum. Dark matters are either self-annihilating or non-self-annihilating. Self-annihilating dark matter can heat the core of the neutron star, consequently increasing the surface temperature and altering the cooling rates of the NS \cite{Pethick92,Page:2004,YAKOVLEV2005,Yakovlev11,Baryakhtar17,Bhat:2019}, while non-self-annihilating dark matter can influence the structural properties of neutron stars \cite{Lavallaz10,CIARCELLUTI201119}; in addition to that, some special WIMPs may collapse the star into a black hole \cite{Goldman98,Kouvaris11}.

In recent research, extensive studies have been conducted to study the effects of dark matter and $\Delta$--isobar on neutron stars (NS). The accretion of dark matter 
has been found to decrease the mass, radius, moment of inertia, binding energy and tidal deformability of NS 
\cite{SANDIN2009,CIARCELLUTI201119,Leung11,LI201270,Ellis18,Tuhin19,Das10,Das2021} while it increases $f-$mode oscillation frequency of NS \cite{Das2021104}. The accretion of the dark matter also affect their curvature \cite{Das:2020}. Their accumulation heated the
star\cite{Kouvaris1082,Baryakhtar17,Bell2018, Acevedo2020}, hence affecting the cooling rate of neutron star. The presence of $\Delta$--isobar affects NS properties by promoting the formation of hyperons at higher densities \cite{Ribes19, Drago14, Dex21, Rather2023}, and softening the EOS, leading to a decrease in mass and radius \cite{torsten10,LI18}, and canonical tidal deformability \cite{Li19}.

In our previous work \cite{Jena}, we have studied the effect of $\Delta-$baryons on the bulk properties of neutron stars. We have found that the inclusion of $\Delta$ softens the EOS in consequence of which the maximum mass, radius and tidal deformability decrease. For a specific value of the $\Delta-$meson coupling constants, $R_{1.4}$ decreases by 1.7 km. In this work, we focus on analyzing the impact of dark matter on $\Delta$ admixed hyperon stars. We chose neutralino as our dark matter candidate, which is a supersymmetric (SUSY) WIMP, which obtains the required relic abundance \cite{drees2018}. We use  NLD \cite{furn87}, IOPB \cite{kumar18}, and G3 \cite{kumar18} parameter sets of the RMF model to investigate the effects of WIMP accretion within the core of hyperon stars with $\Delta-$baryons. We choose WIMPs because they are the most abundant dark-matter particles and are believed to be a relic of the early universe \cite{Griest90, 
 Kolb:1998, Rupin14}. There is a possibility that the WIMPs may have decayed in dense regions of the universe, leading to the production of standard-model particles. 

We calculate the mass, radius, and tidal deformability of hyperon stars for varying amounts of dark matter in the core of the NS . Our results of mass and radius are constrained by observations from NICER (Neutron star internal composition exposure) measurements of PSR J0348+0432 (M $=2.01\pm0.04$ $M_\odot$) \cite{Antoniadis13}, PSR J1614-2230 (M $=1.908\pm0.016 M_\odot$) \cite{Arzoumanian18},PSR J0030+0451 (M $=1.34^{+0.15}_{-0.16} M_\odot$, R $=12.71^{+1.14}_{-1.19}$ km) \cite{Riley19}, PSR J0740+6620 (M $=2.072^{+0.067}_{-0.066}M_\odot$, R $=12.39^{+1.30}_{-0.98}$ km) \cite{Riley21},
PSR J0952-0607 (M $= 2.35\pm0.17 M_\odot$) \cite{Romani22}, PSR J1810+1714: M = $=2.13\pm 0.04 M_\odot$ \cite{Romani_2021} and GW170817 \cite{abbo17,abbo18}. Furthermore, we use the canonical tidal deformability $\Lambda_{1.4} =190^{+390}_{-120}$ \cite{abbo18} that is determined for the EOS
lying within the 90\% credible limit for the GW170817 event detected by LIGO-Virgo collaboration \cite{abbo17}. We also consider the radial constraint of the canonical star from the XMM Newton X-rays and NICER observation: R$_{1.4}=12.45\pm 0.65$ km and R$_{1.4}=11.9\pm 1.22$ km from Ref. \cite{Lattimer_2013}.
We also find correlations between dark matter fermi momentum and bulk properties of the neutron star, namely the canonical radius ($R_{1.4}$), canonical frequency ($f_{1.4}$), canonical tidal deformability ($\Lambda_{1.4}$) and moment of inertia of canonical star ($I_{1.4}$).

This manuscript begins with a brief introduction to dark matter and its influence on neutron star bulk properties, followed by a theoretical framework. A section is devoted to a brief description of our result, and it is ended with a summary and concluding remark.

\section{Theoretical Framework}\label{form1}

\begin{figure}[t]
\vspace{0.5cm}
\begin{center}
\includegraphics[width=1\textwidth]{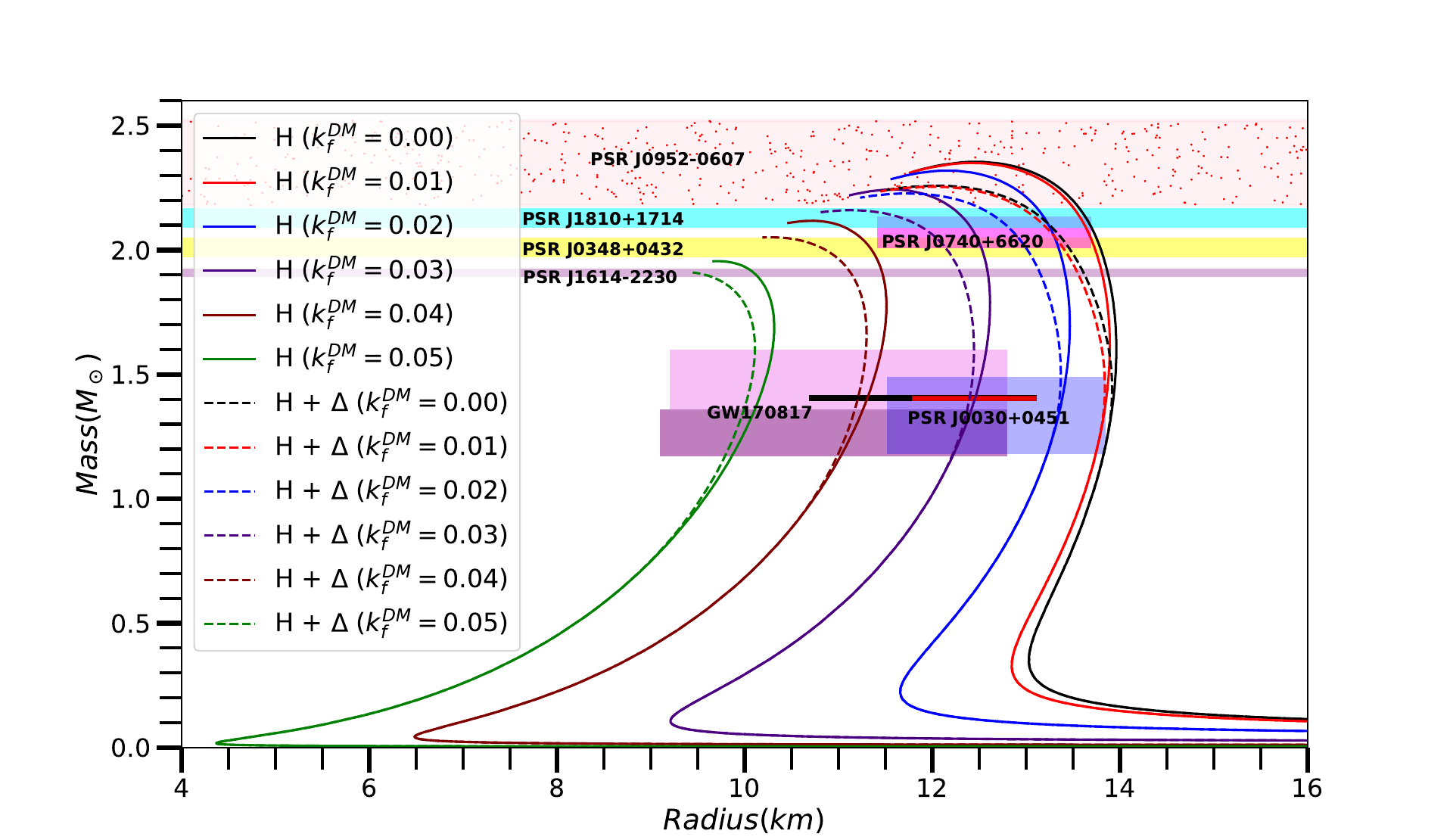}
\end{center}
\caption{M-R profile using NLD parameter set for different value of dark matter Fermi-momentum ($K_f^DM$= 0-0.05 GeV). The H and $\Delta$ represent hyperons and $\Delta-$baryons composition in the core of the NS along with nucleons. The boxes represent the constraints on mass and radius provided by NICER and GW170817. The black and red line represent the constraints on canonical radius as measured in the ref.\cite{Lattimer_2013} and NICER and XMM-Newton X-ray observations, respectively. }
\label{mr}
\end{figure}

\subsection{EOS of nuclear matter}
The equation of state (EOS) of nuclear matter in neutron stars describes the relationship between pressure and energy density or baryon density in the highly dense interior of a neutron star. The EOS is pivotal for studying various properties of neutron stars, including mass, radius, and tidal deformability. It can be modeled using relativistic mean field (RMF) theory, which is a highly effective approach for comprehensively analyzing both finite and infinite nuclear matter \cite{mill72,wale74,rein86,ring96}.

In a nuclear system, baryons interact through the exchange of mesons. The relevant mesons are as follows: isoscalar scalar meson ($\sigma$), isoscalar vector meson ($\omega$), isovector vector meson ($\rho$), and pseudo-scalar meson ($\pi$). RMF theory relies on the mean field approximation, where we only consider mesons with natural parity. Since the $\pi$ meson is a pseudo-scalar meson, its contribution to the mean field approximation is zero. Thus, the RMF framework considers only scalar or vector fields, with isoscalar or isovector fields represented through one-boson exchange theory.

In this context, we focus on the following mesons: the isoscalar scalar meson ($\sigma$), isoscalar vector meson ($\omega$), isovector scalar meson ($\delta$), isovector vector meson ($\rho$), and the strange mesons ($\phi$ and $\sigma^*$). The $\sigma$ meson mediates medium-range attractions between baryons, while the $\omega$ meson facilitates short-range repulsive interactions. The $\sigma^*$ and $\phi$ mesons interact with strange baryons, providing attractive and repulsive forces, respectively. Although the $\delta$ mesons typically have minimal influence on neutron star properties, the study by the ref. \cite{Biswaldelta} have reported a significant impact of $\delta$ mesons on the bulk characteristics of neutron stars. 

The mathematical foundation of RMF theory begins with the effective Lagrangian, which is defined as follows \cite{rein86,mill72,furn87, ring96}:

\begin{eqnarray}
&&{\cal L_N}=\sum_B\overline{\psi}_B\bigg(
i\gamma^{\mu}\partial_{\mu}-m_B+g_{\sigma B}\sigma -g_{\omega B}\gamma_\mu
 \omega^ \mu -\frac{1}{2}g_{\rho B}\gamma_\mu\tau\rho^\mu-g_{\phi_0 B}  \gamma_{\mu} {\phi_0}^{\mu} + g_{\sigma^*B}\sigma^* \bigg)
\psi_B \nonumber \\
&&+ \frac{1}{2}\partial_{\mu}\sigma\partial_{\mu}\sigma 
-m_{\sigma}^2\sigma^2
\left(\frac{1}{2}+\frac{\kappa_3}{3!}\frac{g_{\sigma}\sigma}{m_B}
+\frac{\kappa_4}{4!}\frac{g_{\sigma}^2\sigma^2}{m_B^2}\right)
 - \frac{1}{4}\omega_{\mu\nu}\omega^{\mu\nu} +\frac{1}{2}m_{\omega}^2
\omega_{\mu}\omega^{\mu}\left(1+\eta_1\frac{g_{\sigma}\sigma}{m_B} 
+\frac{\eta_2}{2}\frac{g_{\sigma}^2\sigma^2}{m_B^2}\right) \nonumber \\
&&-\frac{1}{4}R_{\mu\nu}R^{\mu\nu}
+\frac{1}{2}m_{\rho}^2
R_{\mu}R^{\mu}\left(1+\eta_{\rho}
\frac{g_{\sigma}\sigma}{m_B} \right)
+\frac{1}{4!}\zeta_0 \left(g_{\omega}\omega_{\mu}\omega^{\mu}\right)^2 +\frac{1}{2} {m_\phi}^2 {\phi_\mu}{\phi^\mu}+\sum_l\overline{\psi}_l\left(
i\gamma^{\mu}\partial_{\mu}-m_l\right)\psi_l \nonumber \\
&&-\frac{1}{2} {m_\delta}^2 {\delta}^2 +\Lambda_v R_{\mu}R^{\mu}
(\omega_{\mu}\omega^{\mu})+\frac{1}{2}\left(\partial^\mu \sigma^*\partial_\mu\sigma^*-m_\sigma ^{*2}\sigma^{*2}\right),
\end{eqnarray}

where B use for baryons decuplet n, p, $\Lambda$, $\Sigma^-$, 
$\Sigma^0$, $\Sigma^+$, $\Xi^-$, $\Xi^0$ and $\Delta-$baryons 
($\Delta^-, \Delta^0, \Delta^+, \Delta^{++}$), $m_B$ represnts masses of these baryons. The Lagrangian for $\Delta-$baryons comes from the Rarita--Schwinger equation for $\psi_{\Delta}$ field. $m_\sigma, m_\omega, m_\rho, m_\delta, m_\sigma^*$,  and $m_\phi$ are the masses of $\sigma, \omega, \rho, \delta, \sigma^*$ and $\phi$ mesons respectively. $\omega_{\mu\nu}$ and $R_{\mu\nu}$ represent  field tensors
for the $\omega$ and $\rho$ fields, respectively, and are defined as 
$\omega_{\mu\nu}=\partial_\mu \omega_\nu-\partial_\nu \omega_\mu$ and
$R_{\mu\nu}=\partial_\mu R_\nu-\partial_\nu R_\mu$.
\vspace{1em}
\begin{figure}[t]
\vspace{0.5cm}
\begin{center}
\includegraphics[width=1\textwidth]{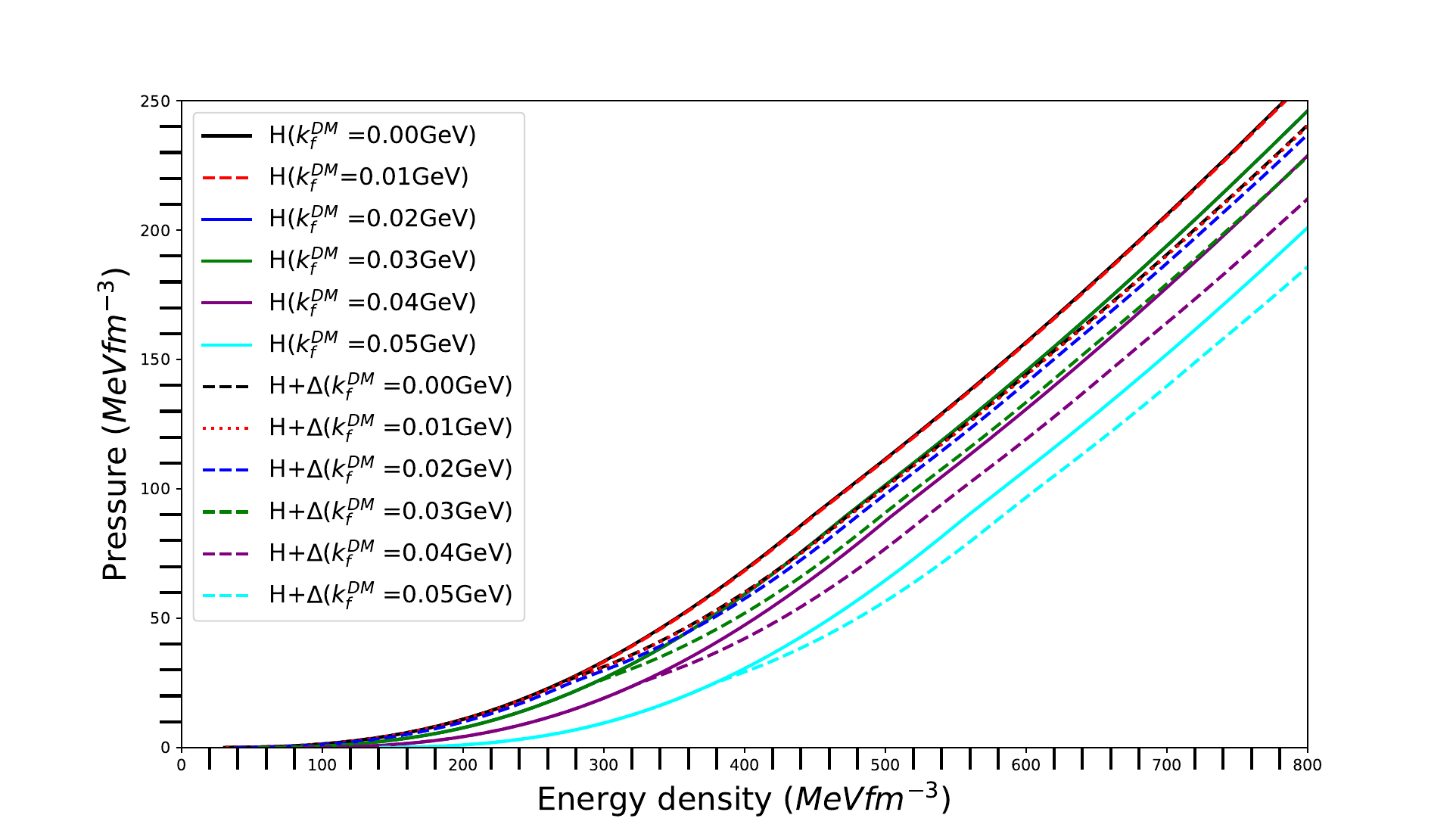}
\end{center}
\caption{Energy density and pressure plot using the NLD parameter set for different values of dark matter Fermi-momentum ($K_f^DM$= 0-0.05 GeV) with and without $\Delta-$baryons. The H and $\Delta$ represent the hyperons and $\Delta-$baryons composition within the 
NS core along with nucleons. }
\label{ep}
\end{figure}

Within $\beta$-stable equilibrium, the conditions that satisfy between chemical potential for the different baryons and leptons
\begin{eqnarray}
    \mu_p = \mu_{\Sigma^+} = \mu_n - \mu_e ,  \mu_n = \mu_{\Sigma^0} = \mu_{\Xi^0} ,
 \cr   \mu_{\Sigma^-} = \mu_{\Xi^-} = \mu_n + \mu_e,
  \cr  \mu_{\mu} = \mu_e,  \mu_{\Delta^-} = 2\mu_n - \mu_p,
  \cr   \mu_{\Delta^0} = \mu_n,  \mu_{\Delta^+} = \mu_p, 
  \cr   \mu_{\Delta^{++}} = \mu_p - \mu_n,
\end{eqnarray}  
and the charge neutrality condition
\begin{eqnarray}
n_p + n_{\Sigma^+} + 2n_{\Delta^{++}} + n_{\Delta^+}  = n_e + n_{\mu^-} + n_{\Sigma^-} + n_{\Xi^-} + n_{\Delta^-}.
\end{eqnarray}

The
energy $\cal E_N$ and pressure $\cal P_N$ densities for nuclear matter can be caculated from 
 energy-momentum tensor $T^{\mu\nu}$ defined as
 \begin{eqnarray}
 T^{\mu\nu} = \sum_{\alpha}\frac{\partial\cal L_N}{\partial 
 (\partial_{\mu}\phi_{\alpha})}\partial^{\nu}\phi_{\alpha}-\eta^{\mu\nu}\cal L_N,       
 \end{eqnarray}
 
\begin{eqnarray}\label{energy}
&&{\cal E_N}=\sum_B\frac{Y_B}{(2\pi)^3}\int_0^{k_F^B} d^3k \sqrt{k ^2+{m_B}^*}
+\frac{1}{8}\zeta_0g_{\omega}^2\omega_0^4 + m_{\sigma}^2\sigma_0^2\left(\frac{1}{2}+\frac{\kappa_3}{3!}
\frac{g_{\sigma}\sigma_0}{m_B}+\frac{\kappa_4}{4!}
\frac{g_{\sigma}^2\sigma_0^2}{m_B^2}\right) \nonumber \\
&& + \frac{1}{2}m_{\omega}^2 \omega_0^2\left(1+\eta_1
\frac{g_{\sigma}\sigma_0}{m_B}+\frac{\eta_2}{2}
\frac{g_{\sigma}^2\sigma_0^2}{m_B^2}\right)  + \frac{1}{2}m_{\rho}^2 \rho_{03}^2\left(1+\eta_{\rho}
\frac{g_{\sigma}\sigma_0}{m_B} \right) +\frac{1}{2}{m_\phi}^2 \phi^2 \nonumber +\frac{1}{2} {m_\delta}^2 {\delta_0}^2 \nonumber\\
&&+\sum_l \int_0^{k_F^l} \sqrt{k^2+{m_l}^2} k^2 dk
+3\Lambda_V \omega_0^2 R_0^2 + \frac{1}{2}m_{\sigma^*}^2\sigma^{*2},
\end{eqnarray}
and
\begin{eqnarray}\label{pressure}
&&{\cal P_N}=\sum_B\frac{Y_B}{3(2\pi)^3}\int_0^{k_F^B}\frac{k^2 d^3k}{\sqrt{k^2+{m_B^*}^2}}
+\frac{1}{24}\zeta_0g_{\omega}^2\omega_0^4
- m_{\sigma}^2\sigma_0^2\left(\frac{1}{2}+\frac{\kappa_3}{3!}
\frac{g_{\sigma}\sigma_0}{m_B}+\frac{\kappa_4}{4!}
\frac{g_{\sigma}^2\sigma_0^2}{m_B^2}\right) \nonumber \\
&& + \frac{1}{2}m_{\omega}^2 \omega_0^2\left(1+\eta_1
\frac{g_{\sigma}\sigma_0}{m_B}+\frac{\eta_2}{2}
\frac{g_{\sigma}^2\sigma_0^2}{m_B^2}\right)  + \frac{1}{2}m_{\rho}^2 \rho_{03}^2\left(1+\eta_{\rho}
\frac{g_{\sigma}\sigma_0}{m_B} \right)-\frac{1}{2} {m_\delta}^2 {\delta_0}^2 \nonumber \\
&&+\Lambda_V R_0^2 \omega_0^2 - \frac{1}{2}m_{\sigma^*}^2\sigma^{*2} +\frac{1}{3\pi^2}\sum_l \int_0^{k_F^l} \frac{k^4 dk}{\sqrt{k^2+m_l^2}},
\end{eqnarray}
where $l$ stands for the leptons and $m_B^*$ is the effective masses of baryons which is defined below.
\begin{eqnarray}
    m_B^* = m_B - g_{\sigma B} \sigma_0 - g_{\delta B} \tau_B \delta_0 -g_{\sigma^*B} \sigma^*
\end{eqnarray}

\subsection{EOS of dark matters}
Here, we consider the neutralino, which is a fermionic candidate of dark matter. The mass of the neutralino is 200 GeV, and it interacts 
with baryons through the standard model (SM) Higgs boson. The mass of the Higgs boson is 125 GeV, and the coupling constant (y) between dark matter and the SM 
Higgs boson has a value ranging from 0.001 to 0.1. For our specific case, we use y=0.07.
The Lagrangian for dark matter interaction with baryons is given by  \cite{Panotopoulos17,Tuhin19,Quddus2020,Das10,Das2021}
\begin{eqnarray}\label{LD}
    \cal L_{DM} = \overline{X} [\textit{i} \gamma^\mu \partial_\mu - M_X + \textit{yh}]X
&&+\frac{1}{2} \partial_\mu \textit{h} \partial^\mu \textit{h} 
    -\frac{1}{2}M_{\textit{h}}^2h^2 + \sum_B f_B \frac{m_B}{v}\overline{\psi}_B \textit{h}\psi_B   
\end{eqnarray}\label{LD1}
where \textit{v} is the vacuum expectation value.

The last part of Eqn.(\ref{LD}) denotes the Yukawa interaction, which represents the direct interaction between the Higgs field and baryons. In this equation, 
\textit{f} refers to the proton-Higgs form factor, with a specific value of 0.35 chosen based on lattice calculations \cite{cline13}.
From Euler-Lagrange equation we can find the following equation of states for dark matter and baryon interaction: 
\begin{eqnarray}
  &&  (i\gamma^\mu \partial_\mu - \mathcal{M}_\chi + yh)\chi = 0, \nonumber \\
   && \partial_\mu \partial^\mu h + M_h^2 h= y \overline{\chi}\chi + \sum_B \frac{f_B m_B}{v} \overline{\psi}_B\psi_B
\end{eqnarray}

Using RMF approach, we get
\begin{eqnarray}
    \textit{h}_0 = \frac{\textit{y}\mathcal{\bigg <\overline{X}X\bigg>} + \sum_B \frac{f_B m_B}{v}\overline{\psi}_B\psi_B}{M_\textit{h}^2},\nonumber\\
    (\textit{i}\gamma^\mu\partial_\mu - M_{\mathcal{X}}^*)\mathcal{X} = 0,
\end{eqnarray}

where $M_{\mathcal{X}}^*$ is the effective mass of dark matter which is defined as
\begin{eqnarray}
    M_{\mathcal{X}}^* = M_\mathcal{X} - \textit{y} h_0
\end{eqnarray}
and the dark matter scalar density $\rho_s^{DM} = \mathcal{\bigg <\overline{X}X\bigg>}$ is
\begin{eqnarray}
    \rho_s^{DM} = \frac{\gamma}{2\pi^2} \int_0^{k_f^{DM}} \frac{M_{\mathcal{X}}^*}{\sqrt{ (M_{\mathcal{X}}^*)^2 + k^2}}dk.
\end{eqnarray}

The dark matter energy density and pressure can be calculated from energy-momentum tensor as
\begin{eqnarray}
    \mathcal{E_{DM}} = \frac{1}{\pi^2}\int_0^{k_f^{DM}} k^2 \sqrt{k^2 + (M_{\mathcal{X}}^*)^2}dk + \frac{1}{2} M_\textit{h}^2 \textit{h}_0^2,
  \end{eqnarray}  
  and
  \begin{eqnarray}
    \mathcal{P_{DM}} = \frac{1}{3\pi^2} \int_0^{k_f^{DM}} \frac{k^4}{\sqrt{k^2 + (M_{\mathcal{X}}^*)^2}} dk - \frac{1}{2} M_\textit{h}^2 \textit{h}_0^2.
\end{eqnarray}
Hence, the total Lagrangian can be written as
\begin{eqnarray}
    \cal L = L_N + L_{DM}.\nonumber
\end{eqnarray}
Similarly, the total energy and pressure are
\begin{eqnarray}
    \cal E = E_N + E_{DM}, P = P_N + P_{DM}
\end{eqnarray}
\begin{figure}[t]
\vspace{0.5cm}
\begin{center}
\includegraphics[width=1\textwidth]{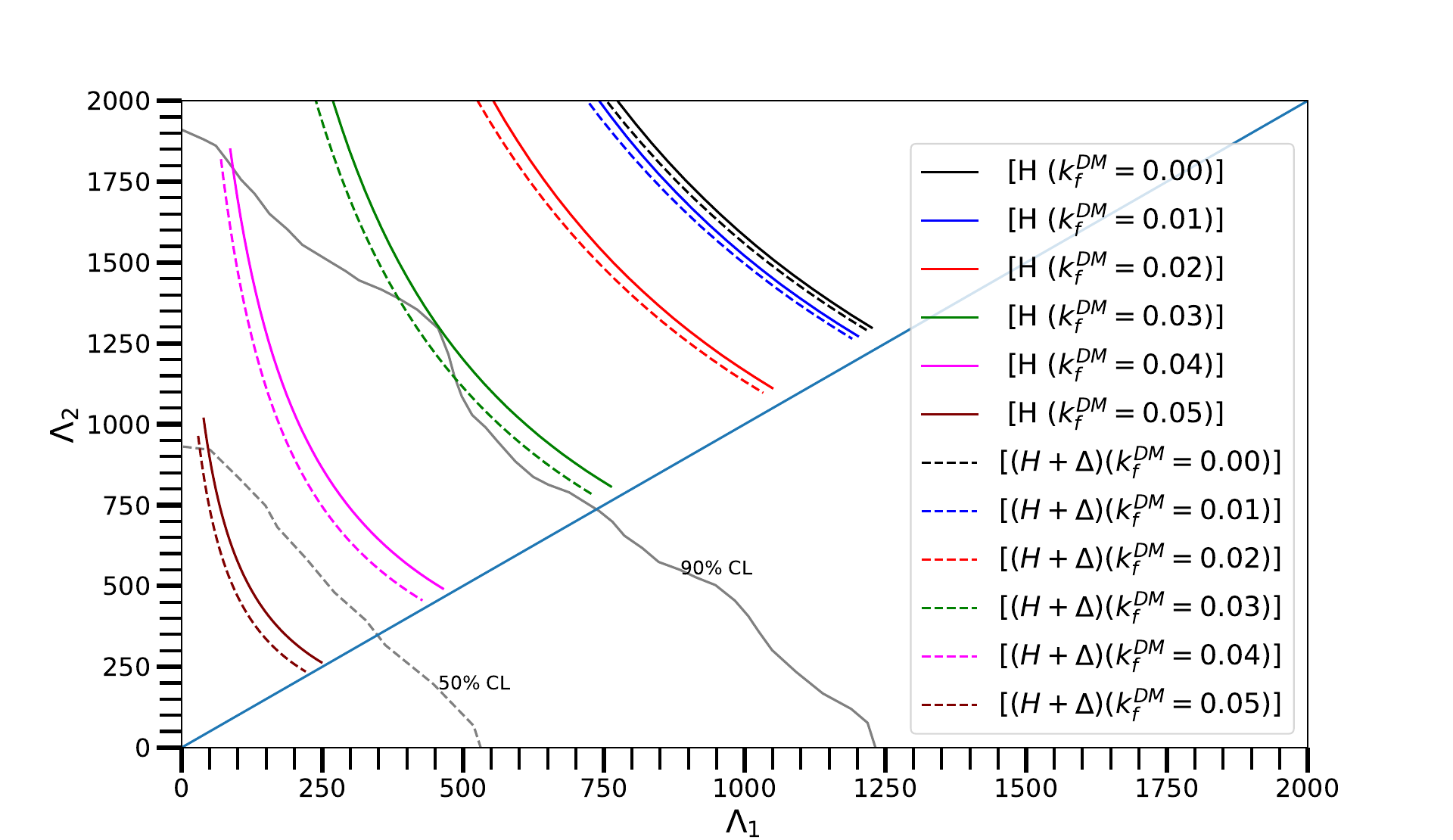}
\end{center}
\caption{Tidal deformability  with NLD parameter set for different value of dark matter Fermi-momentum ($K_f^DM$= 0-0.05 GeV) with and without $\Delta$-baryons. Here, H and $\Delta$ represent hyperons and $\Delta-$baryons in the hyperon star.}
\label{td}
\end{figure}
\subsection{Calculation of mass and radius}
A non-rotating neutron star is assumed to have a spherical shape in its equilibrium state and metric can be written as
\begin{eqnarray}\label{metric}
ds^2=-e^{2\Phi(r)} dt^2+e^{2\Lambda(r)}dr^2+r^2d\theta^2+r^2 sin^2\theta d\phi^2,
\end{eqnarray}
where $\Lambda (r)$ and $\Phi(r)$ are the metric functions. The solution of Eqn.(\ref{metric}) gives rise to  Tolman-Oppenheimer-Volkoff (TOV) equation \cite{tolm39,oppe39,glenb97}
\begin{eqnarray}
\frac{\partial m(r)}{\partial r}=4\pi r^2 \epsilon(r),
\end{eqnarray}

\begin{eqnarray}
\frac{\partial P(r)}{\partial r}= - \frac{(P(r)+\epsilon(r))(m(r)+4\pi r^3 P(r))}{r(r-2m(r))},
\end{eqnarray}
and 

\begin{eqnarray}
    \frac{d\Phi(r)}{dr} = \frac{(m(r)+4\pi r^3 P)}{r(r-2m(r))},
\end{eqnarray}

where m(r) represents the mass enclosed within the radius r. P(r) and $\epsilon(r)$ represent the total pressure and total energy density, respectively. 
The tolal energy and the total pressure can be used to calculate the mass and radius of the hyperon star with dark matter from the TOV equation.
\subsection{Calculation of $f-$mode oscillation frequencies}
The oscillation of neutron stars can be initiated by various factors, such as binary interaction, mass accretion from a companion star, nutation of the star, rapid 
rotation, and magnetic field. Neutron stars can exhibit different types of oscillation modes: \textit{p}-mode (pressure modes), \textit{g}-mode (gravity modes), 
\textit{f}-mode (fundamental modes), \textit{s}-mode (shear modes), \textit{t}-modes (toroidal modes), \textit{r}-modes (Rossby modes), and \textit{i}-modes 
(interfacial modes) due to different forces acting on a displaced mass element, such as gravity, pressure gradients, elastic forces, magnetic fields, and 
centrifugal and coriolis forces in rotating neutron stars. 

In this work, we examine the non-radial oscillation of the dark matter admixed hyperon stars and compute the $f-$mode frequencies for different compositions of dark 
matter in the hyperon star using Cowling approximation (i.e., neglecting the metric perturbation) \cite{Cowling41}. To do this, we solve the following 
differential equations\cite{sotani11,C14,Ignacio18}:
\begin{eqnarray}\label{frequency}
    \frac{dW(r)}{dr}=\frac{d\varepsilon}{dP}[\omega^2 r^2 
    &&e^{\Lambda(r)-2\Phi(r)}V(r) + \frac{d\Phi(r)}{dr} W(r)]\nonumber\\ 
    &&- l(l+1)e^{\Lambda(r)} V(r)
\end{eqnarray}\label{frequency1}
and
 \begin{eqnarray}
     \frac{dV(r)}{r} = 2\frac{\Phi(r)}{r} V(r) - \frac{1}{r^2} &&e^{\Lambda(r)} W(r),
 \end{eqnarray}
 
 The Lagrangian displacemant vector $\eta$ of a perturbed fluid is defined as 
\begin{equation}
    \eta = \frac{1}{r^2} (e^{-\Lambda(r)} W(r))- V(r)\partial_\theta - \frac{V(r)}{sin^2\theta}\partial_\phi) Y_{lm}(\theta,\phi),
\end{equation}
where Y$_{lm}$($\theta,\phi$) repesent the spherical harmonic.

The eigenvalues of frequency can be found by solving the Eqn.(\ref{frequency})
with the boundary conditions 
\begin{eqnarray}
W(r) = Br^{l+1}, V(r) = -\frac{B}{l}r^l,
\end{eqnarray}
and 
\begin{eqnarray}
\omega^2e^{\Lambda(R)-2\Phi(R)}V(R) + \frac{1}{R^2} \frac{d\Phi(r)}{dr}|_{r=R} W(R) = 0.
\end{eqnarray}

\subsection{Calculation of tidal deformability}
Tidal deformability is one of the key observable properties of neutron stars within binary systems. In such system of two stars, the primary star and its companion star, orbit around each other. As they move in their orbital paths, each star experiences a gravitational pull from the other, known as the tidal force. This force becomes stronger as the stars get closer to merging. During the merger, the stars emit gravitational waves that contain crucial information about their characteristics. The study of tidal deformability gained significant attention after the detection of gravitational waves resulting from a binary neutron star (BNS) merger. This particular event, referred to as GW170817, occurred on August 17, 2017, and was detected by the LIGO and Virgo collaborations. The GW170817 event also provides important constraints on tidal deformability i.e $\Lambda_{1.4} =190^{+390}_{-120}$ \cite{abbo18}.

The tidal deformability of a neutron star (NS) tells how much the NS can be deform by the gravitational pull of its companion star. This deformability is affected by the EOS and internal compostion of neutron star. The tidal deformability correlates with the quadrupole moment $Q_{ij}$ through the formula 
\begin{eqnarray}\label{quadr}
    \lambda = \frac{Q_{ij}}{\epsilon_{ij}},
\end{eqnarray}\label{quadr1}
 where $\epsilon_{ij}$ represents the perturbed tidal field causing the deformation. 
The tidal deformability can also be represent through tidal love number $k_2$ \cite{Hinderer08} by the expression
\cite{hind10, Raithel18,Chatziioannou2020,Leung22}
\begin{equation}
\lambda =\frac{2k_2 R^5}{3},\
\end{equation}

and the dimension less tidal deformability is
\begin{equation}
\Lambda = \frac{2k_2R^5}{3M^5}=\frac{2k_2 }{3C^5}.\
\end{equation}
M and R represent mass and radius of NS where C=M/R is the compactness of the NS.
\begin{figure}[t]
\begin{center}
\includegraphics[width=1\textwidth]{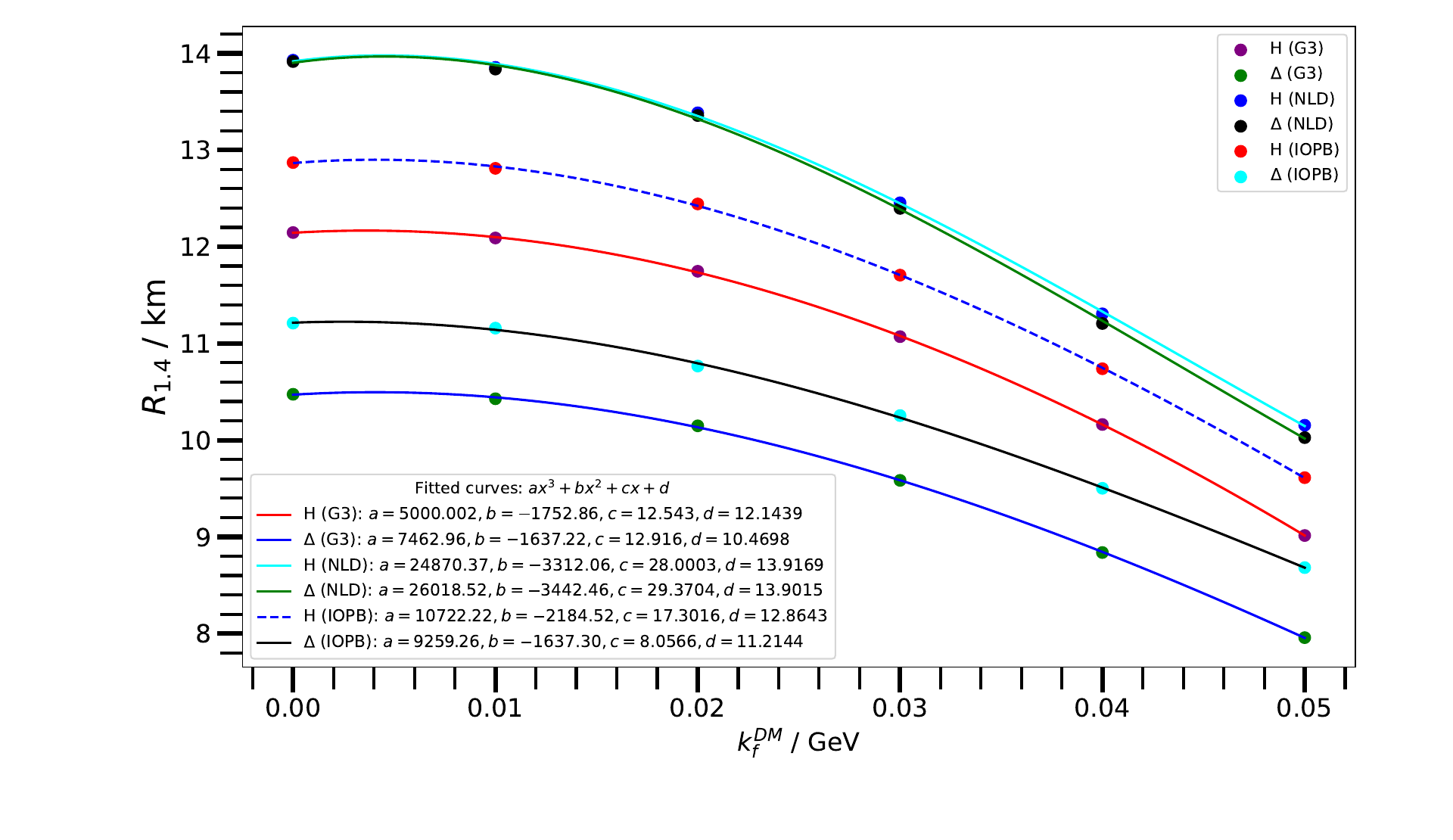}
\end{center}
\caption{ Canonical radius ($R_{1.4}$) for different values of dark matter Fermi-momentum ($K_f^{DM}$= 0-0.05 GeV) using G3, NLD, IOPB parameter sets of RMF model. The H represent the neutron star core composed of hyperons and $\Delta$ represent hyperons with $\Delta-$baryons composition. The x represents $k_f^{DM}$.}
\label{r14}
\end{figure}
\subsection{Setting coupling constant for hyperons and $\Delta-$baryons}
The coupling constant for nucleons (n, p) are fixed to ground state property of nuclear matter and finite nuclei. Set of these coupling constant are named as NLD, IOPB, and G3. But as we know there are not much experimental data available for hyper nuclei, so the hyperon coupling constants are mostly predicted by theory. 
 The coupling constants for different hyperons and $\Delta-$baryons are set using the following equation:
\begin{eqnarray}
U_{Y} = {m_n}(\frac{m_n^*}{m_n}-1)X_{\sigma Y}+(\frac{g_\omega}
{m_\omega})^2\rho_0 X_{\omega Y},
\end{eqnarray}

where Y stands for the different hyperons 
($\Lambda, \Sigma, \Xi $) and $\Delta-$baryons ($\Delta^-, \Delta^{0}, \Delta^+, \Delta^{++}$). 
$X_{\sigma Y}$ and $X_{\omega Y}$ are the coupling constants of the
hyperon-meson interactions, and $\rho_0$ is the saturation density.
The coupling constant for $\Lambda, \Sigma$ and $\Xi$ hyperons are set as ${U_{\Lambda}} = -28 MeV$ \cite{batt97,mill88},
${U_{\Sigma}} = +30$ MeV \cite{kohn06,hara05,frie07},
 and ${U_{\Xi}}= -18 $ MeV \cite{glen91,scha00}.
The interaction strength of
hyperon and $\rho$ meson interaction \cite{dove84,scha94}, hyperon and $\phi$-meson interaction are set by using SU(6) symmetry
$X_{\Lambda\rho}=0$, $X_{\Sigma\rho}=2$, $X_{\Xi\rho}=1$,
$X_{\phi\Lambda}=- \frac{\sqrt{2}}{3}g_{\omega N}$,  $X_{\phi\Sigma}=
-\frac{{\sqrt{2}}}{{3}}$$g_{\omega N}$, $X_{\phi\Xi}=-\frac{2\sqrt{2}}{3}$$g_{\omega N}$.

The $\Delta$ coupling constant can be chosen using the conditions, -150 MeV$\leq U_{\Delta} \leq$ -50 MeV and $0\leq X_{\sigma\Delta} - X_{\rho_\Delta}\leq 0.2$ \cite{WEHRB89,Drago14}.
Based on this condition, the $X_{\sigma\Delta}$, $X_{\omega\Delta}$ and $X_{\rho\Delta}$ are set as 1.2, 1 and 1 respectively. 
\section{Result and Discussion}
\subsection{Effect of dark matter on neutron star bulk properties:}
 The baryonic matter present in the NS core balances gravity through nuclear pressure. This nuclear pressure arises from the nucleon degeneracy pressure and strong interactions. Weakly interacting dark matter particle accreted into the core of the NS due to its high gravitational mass. The dark matter particle interact very weakly with the nuclear mattter present in the core of the NS. Baryon mostly contribute to the pressure of EOS however the contibution of the DM to the pressure is  comparatively less due to its weak interaction with nuclear matter. So, the EOS with dark matter become softer leading to  decrease in mass and radius. We change the DM content inside the neutron star by changing its Fermi momentum from 0-0.05 GeV. Our study also informs about the change in mass-radius profile of a $\Delta-$admixed hyperon star with DM content. Calculating the EOS of the $\Delta-$admixed hyperon star, we take the $\Delta-$baryon mass as 1232 MeV and the coupling constant set to $X_{\sigma\Delta}=1.1$, $X_{\omega\Delta}=1.0$ and $X_{\rho\Delta}=1.0$. We use the NLD (non--linear derivative) parameter set of the RMF model for the calculation of neutron star bulk properties in the presence of dark matter. The NLD parameter set has high incompressibility, which
provides a stiff EOS, and the maximum mass reaches to the highest observed mass of PSR J0952-0607 \cite{Romani22}. In addition to this, the NLD parameter set satisfies most of the NICER constraints \cite{Jena}. The presence of hyperons and $\Delta$--baryons makes the EOS softer and reduces the maximum possible mass and radius. The inclusion of dark matter further softens the EOS, and due to this, some of the parameter sets are not be able to reach up to the maximum observed mass of PSR J0952-0607 \cite{Romani22}. Some of the parameter sets are not able to reproduce the canonical mass, which is needed to study of properties of the canonical star. The NLD parameter set is appropriate for studying the DM effects on $\Delta-$admixed hyperon star. Further we take IOPB, G3 parameter sets whose maximum masses lie above the canonical mass and are helpful to study the correlation among the properties of the canonical star.

From Fig. \ref{mr} and Table \ref{tab1}, we find that the population of $\Delta$--baryons in the core of the hyperon star reduces the maximum mass and the canonical radius $R_{1.4}$. Adding dark matter to the EOS, the accretion of dark matter inside the NS's core leads to a more significant decrease in maximum mass and radius. There is a significant change in these bulk properties with the increasing values of dark matter Fermi momenta. When we increase the dark matter Fermi momentum, the dark matter composition inside the core increases, which adds gravitational weight but not that much pressure. With an increasing value of $k_f^{DM}$ the energy density increases faster than the pressure, causing a softer EOS (see Fig. \ref{ep}). Due to the softness of the EOS, the maximum mass and radius decrease. From Table.\ref{tab1}, we can see that the canonical radius for a hyperon star composed of baryonic matter only decreases by 3.774 km, 3.258 km and 3.132 km for NLD, IOPB and G3 parameter sets, respectively, with the dark matter Fermi momentum $k_f^{DM}$ = 0.05 GeV. For $\Delta-$admixed hyperon star the changes are 3.888 km, 2.254 km and 2.516 km for NLD, IOPB, and G3, respectively (see Table.\ref{tab1}). The decrease in $R_{1.4}$ is due to the more compact structure of the star. Since the DM particle does not provide any strong repulsive forces, unlike baryonic matter, the presence of DM leads to strong gravitationally bound system; the star becomes more compact, causing a smaller canonical radius. In addition to that for $k_f^{DM}$ = 0.03 GeV, the NLD parameter satisfies all the NICER and GW170817 constraints both for hyperon stars with and without $\Delta$--admixture. 

For detailed study, we plot the energy--pressure with different dark--matter Fermi momentum as shown in Fig. \ref{ep}. With an increasing value of dark matter Fermi momentum, the EOS becomes softer, which leads to a decrease in maximum mass and canonical radius. 
Furthermore, we study how the dark matter Fermi momenta affect the tidal deformability of the hyperon stars. We plot the tidal deformability of a hyperon star with and without $\Delta$--admixture shown in Fig. \ref{td}. Calculation suggest that tidal deformability for NLD parameter using dark matter Fermi momenta $k_f^{DM}$= 0.04 and 0.05 GeV lies within the 90\%  credible limit of the GW170817 event. The value of tidal deformability for $k_f^{DM}$= 0.04 and 0.05 GeV lies within the range 70--580. The value of tidal deformabilities of the NLD parameter set for $k_f^{DM}=$ 0.04 and 0.05 GeV are 371.858 and 192.538 respectively. Analyzing Fig. \ref{mr} and Fig. \ref{td} we find that the hyperonic star with dark matter Fermi momenta $k_f^{DM}$ = 0.04 and 0.05 GeV satisfies the GW170817 constraints for the NLD parameter set.

\begin{figure}[t]
\begin{center}
\includegraphics[width=1\textwidth]{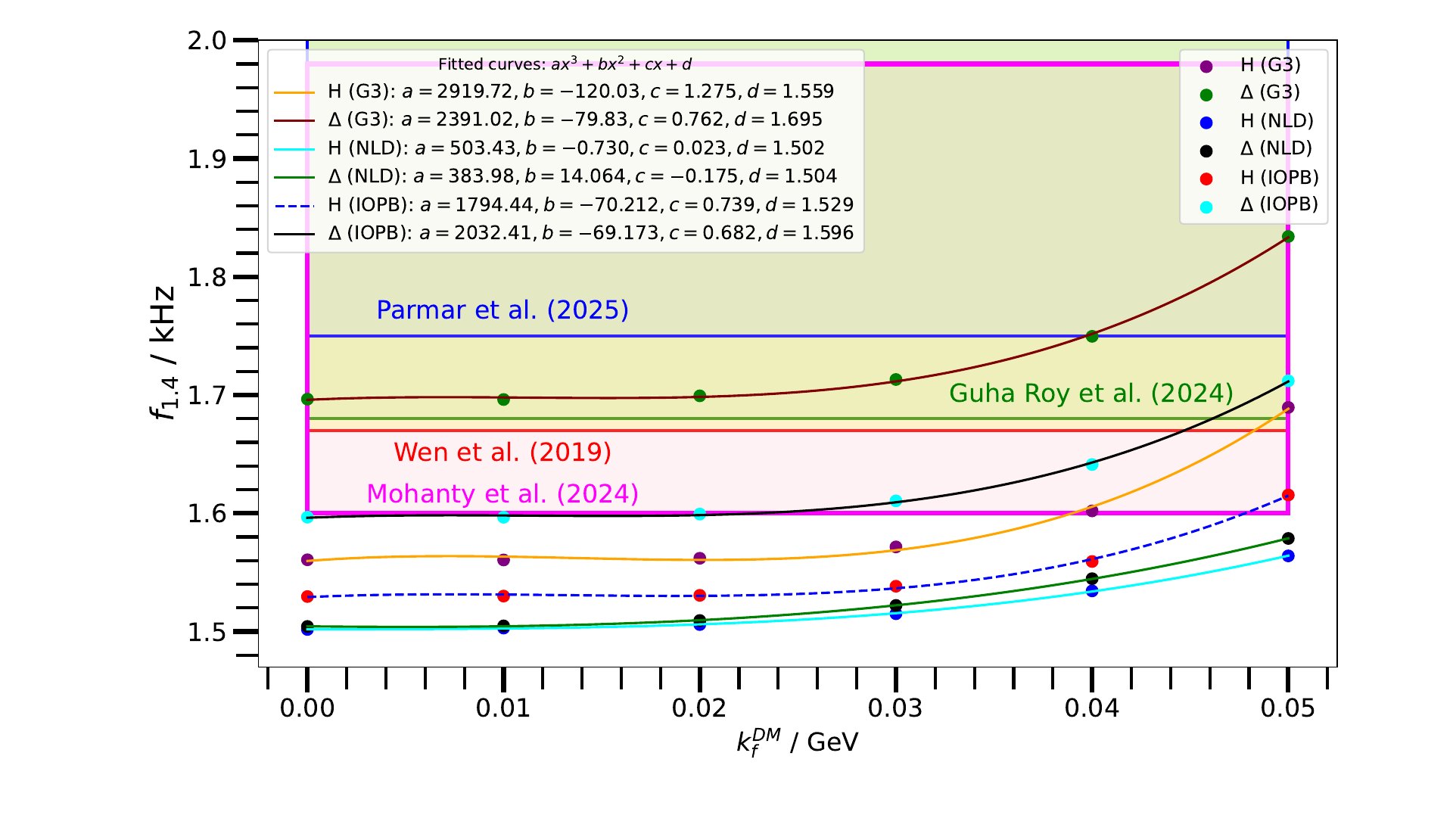}
\end{center}
\caption{Canonical frequency $f_{1.4}$ of the NS for different value of dark matter Fermi-momentum ($k_f^{DM}$= 0-0.05 GeV) using NLD, G3 and IOPB parameter sets of RMF model. The H represent the neutron star core composed of hyperons and $\Delta$ represent hyperons with $\Delta-$baryons composition. The x represents $k_f^{DM}$.}
\label{f14}
\end{figure}
\subsection{Correlation between the dark matter fermi momenta and bulk properties of the neutron star}
In this section, we study how the DM Fermi momenta are correlated with the bulk properties of the neutron star. Here, we use the NLD, G3, and IOPB parameter sets of the RMF model to study the correlations. We use these parameter sets as their maximum masses lie above the canonical mass (1.4 $M_\odot$) and to inquiry whether these correlations are universal or parameter-dependent. We have found a cubical correlation between the dark matter Fermi momenta and various bulk properties of the neutron star. 
These correlations are described in detail below.
\subsubsection{Correlation between canonical radius and dark matter Fermi momenta ($R_{1.4}=$$f(k_f^{DM})$):}
In this section, we examine the relationship between the canonical radius and the Fermi momentum of dark matter. Figure \ref{r14} illustrates how \( R_{1.4} \) varies with different values of \( k_f^{DM} \) using various Relativistic Mean Field (RMF) parameter sets, namely G3, IOPB, and NLD. The figure clearly indicates that each RMF parameter set exhibits a non-linear decreasing trend, which suggests that \( R_{1.4} \) decreases as the value of \( k_f^{DM} \) increases. This decreasing trend demonstrates a correlation between \( R_{1.4} \) and \( k_f^{DM} \), which is well-represented by a third-order polynomial function defined as:

\begin{equation}
    R_{1.4} = a (k_f^{DM})^3 + b (k_f^{DM})^2 + c (k_f^{DM}) + d.
\end{equation}

Here, the correlation coefficients \( a \), \( b \), \( c \), and \( d \) vary with different parameter sets. We also investigate this correlation for various compositions of neutron star core, focusing specifically on hyperons with and without \(\Delta-\)baryons. Our findings indicate that the values of \( a \), \( b \), \( c \), and \( d \) are influenced by the composition of the star's core but the functional form remains same. We have also taken different RMF parameter sets and observe the same functional form for these parameter sets. From all these observations, we conclude that $R_{1.4}$ and $k_f^{DM}$ exhibit a cubic correlation, and this correlation strongly depends on the parameter sets and the composition of the neutron star.

\begin{figure}[t]
\begin{center}
\includegraphics[width=1\textwidth]{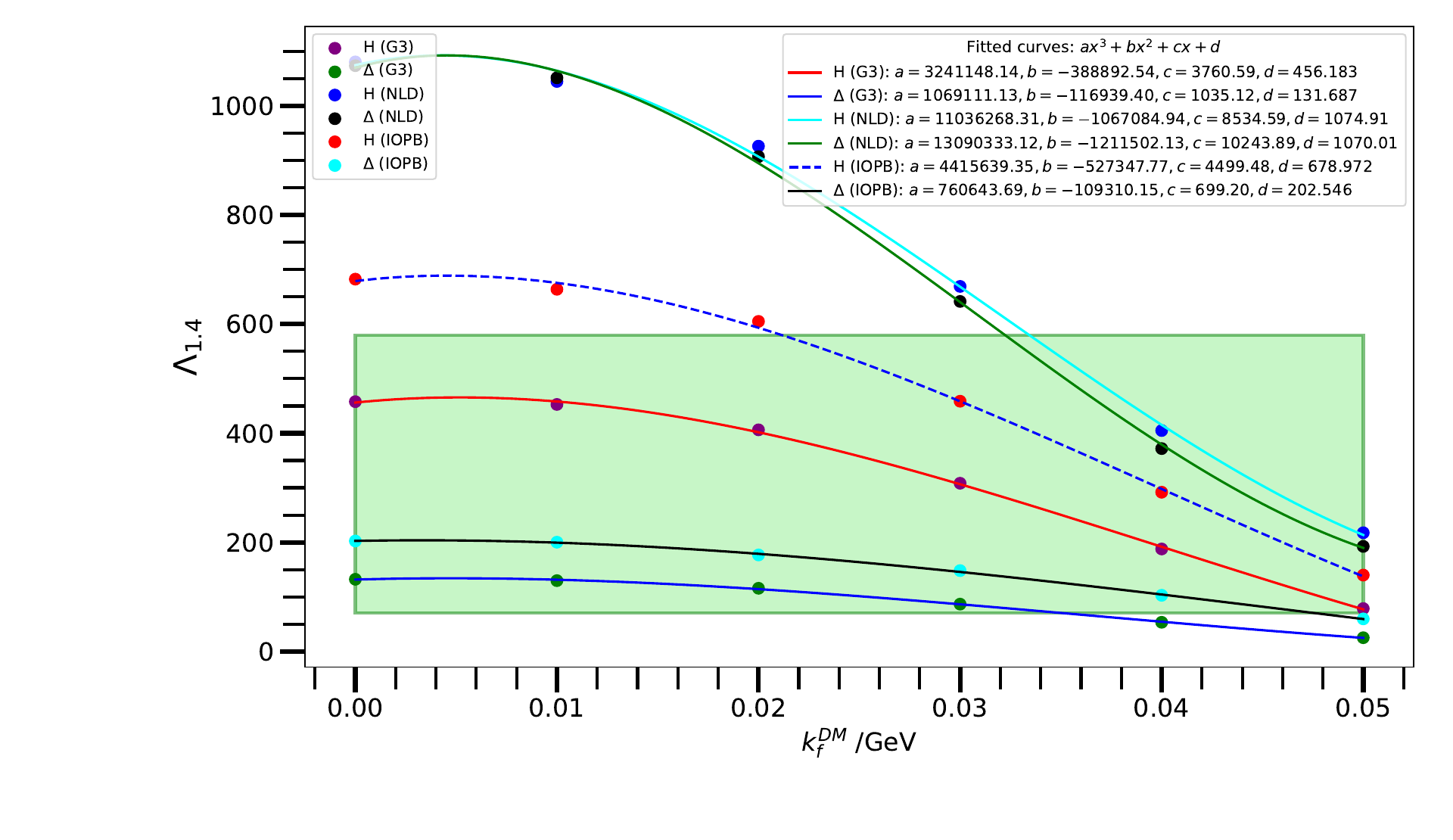}
\end{center}
\caption{ Canonical tidal deformability ($\Lambda_{1.4}$) with NLD, G3, and IOPB parameter sets for different values of dark matter Fermi momentum ($K_f^DM$= 0-0.05 GeV). The H represent the hyperon star without $\Delta-$baryons composition and $\Delta$ represent hyperon star with $\Delta-$baryons composition. The x represents $k_f^{DM}$.}
\label{l14}
\end{figure}
\subsubsection{Correlation between canonical frequency and dark matter Fermi momenta ($f_{1.4}=$$f(k_f^{DM})$):}
In this section, we study how the dark matter Fermi momenta affect the $f-$mode frequency of neutron star. From Fig. \ref{f14} and Table \ref{tab1}, we find that for each parameter set, the $f-$mode frequency of the canonical star ($f_{1.4}$) increases with rise in the value of $k_f^{DM}$. This is due to the increase compactness of neutron star as $f-$mode frequency directly related to the compactness of the neutron star. For same mass the radius of the star reduces for which compactness also increases. Furthermore, in Fig. \ref{f14}, we have shown that the canonical frequency is correlated with the
dark matter Fermi momenta $k_f^{DM}$ by a third order polynomial equation defined as: 

\begin{equation}
    f_{1.4} = a (k_f^{DM})^3 + b (k_f^{DM})^2 + c (k_f^{DM}) + d.
\end{equation}
To investigate the dependence of this correlation on various parameters and compositions of the core, we utilise the NLD, G3, and IOPB parameter sets, considering the composition of the hyperon star both with and without $\Delta-$baryons.The functional form of this correlation is same for all parameter sets with and without $\Delta-$baryons. However, the correlation coefficients $a,$ $b,$ $c,$ and $d$ values depend on the parameter sets and compositions of the core. 

We constraint the canonical frequency with the constraints proposed by Wen et al. (2019) \cite{Wen19}: $f_{1.4}=1.67-2.18$ kHz, Guha Roy et al. (2024) \cite{GuhaRoy_2024}: $f_{1.4}=1.8^{+0.7}_{-0.12}$, Mohanty et al. (2024) \cite{mohanty2024}: $f_{1.4}=1.75^{+0.23}_{-0.15}$ and  Parmer et al. (2025) \cite{Parmar_2025} : $f_{1.4}=1.97^{+0.17}_{-0.22}$ kHz. From Fig. \ref{f14}, we find the G3 parameter set with $\Delta-$admixed hyperon composition satisfies all the constraints with DM content $k_f^{DM}$ more than 0.04GeV while NLD parameter set does not satisfy any constraint of $f_{1.4}$ with DM content. The other parameter set IOPB does not satisfy the $f_{1.4}$ constraints with a lower value of $k_f^{DM}$, it approaches the constraints when the DM content increases. 

\subsubsection{Correlation between caonical tidal deformability and dark matter Fermi momenta (\(\Lambda_{1.4}=\) \(f(k_f^{DM})\)):}
Fig. \ref{l14} illustrates how the canonical tidal deformability varies with the Fermi momenta of dark matter. From Table \ref{tab1} and Fig. \ref{l14}, we observe that the tidal deformability of the IOPB parameter meets the GW170817 constraint, both with and without the inclusion of dark matter. The G3 parameter set falls within the range of the tidal deformability constraint specified by GW170817 for dark matter Fermi momenta from 0.00 to 0.03 GeV. Meanwhile, the NLD parameter set satisfies this constraint for dark matter Fermi momenta of 0.04 and 0.05 GeV. In addition to that in Fig. \ref{l14}, we fit the curves using polynomial equations and discovered a third-order polynomial correlation between tidal deformability and Fermi momenta. This correlation is defined as:
\begin{equation}
    \Lambda_{1.4} = a (k_f^{DM})^3 + b (k_f^{DM})^2 + c (k_f^{DM}) + d.
\end{equation}
The correlation coefficients \( a \), \( b \), \( c \), and \( d \) vary for the composition and parameter sets, indicating a strong dependence of this correlation on the parameter set and the composition of the neutron star core. However, the functional form remains same for different parameter set and different compositions. In Fig. \ref{l14}, we constrain $k_f^{DM}$ with the help of the constraint imposed by GW170817 ($\Lambda_{1.4} =190^{+390}_{-120}$) \cite{abbo18}. The G3 (with hyperons) and IOPB (with hyperon + $\Delta$) parameter sets satisfy the GW170817 constraint for $k_f^{DM}=0.00-0.05$ GeV. The NLD (both for hyperon and $\Delta-$admixed hyperon) and IOPB (with hyperon) parameter sets without DM lie above the GW170817 constraint. However, with DM for $k_f^{DM}=0.04$ and 0.05 GeV, these parameters satisfy this constraint. Comparing the result with the GW170817 constraint, we can constrain $k_f^{DM}$ to 0.04 and 0.05 GeV. 

\begin{figure}[t]
\begin{center}
\includegraphics[width=1\textwidth]{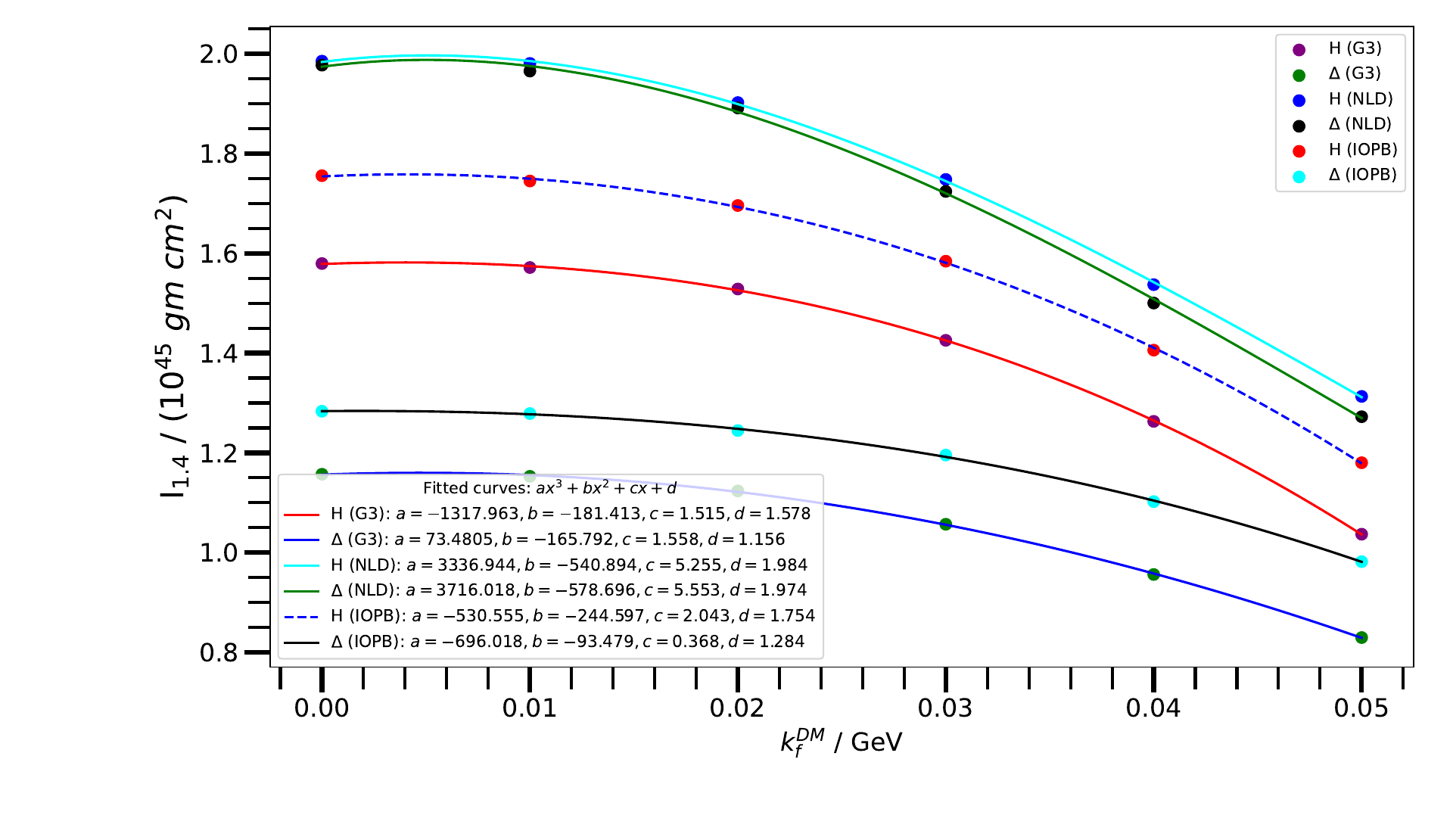}
\end{center}
\caption{ Moment of inertia of canonical star ($I_{1.4}$) with NLD, G3, and IOPB parameter set for different values of dark matter Fermi momentum ($K_f^DM$= 0-0.05 GeV). The H represent the hyperon star without $\Delta-$baryons composition and $\Delta$ represent hyperon star with $\Delta-$baryons composition. The x represents $k_f^{DM}$.}
\label{i14n}
\end{figure}

\begin{sidewaystable}
\vspace{2.0 cm}
\centering
\caption{
	Maximum mass (M), corresponding radius (R), canonical radius and canonical tidal deformability, canonical frequency, and canonical moment of inertia of hyperon star with and without $\Delta$--isobars in various parameter sets. Here, we take $\Delta$--coupling constant: $X_{\sigma \Delta}$ = 1.1, $X_{\omega \Delta}$ = 1 and $X_{\rho \Delta}$ = 1.\\
 }
\renewcommand{\tabcolsep}{0.20 cm}

{\begin{tabular}{|c| c| c| c| c| c| c| c| c| c| c|c|c|c|c|c| }
\hline 

&&\multicolumn{6}{|c|}{Hyperon star
without $\Delta$}
&\multicolumn{6}{|c|}{Hyperon star with $\Delta$}\\

\hline
&&&&&&&&&&&&&\\[5pt]
parameter&$k_f^{DM}$ &M &R  & $R_{1.4}$  & $\Lambda_{1.4}$ & $f_{1.4}$ & $I_{1.4}$ / $10^{45}$
	& M &R &$R_{1.4}$ & $\Lambda_{1.4}$ & $f_{1.4}$ & $I_{1.4}$ / $10^{45}$\\
	sets &GeV& ($M_\odot$) &(km)& (km)   &&KHz&(gm cm$^2$)
	& ($M_\odot$)&(km)& (km)  &&KHz& (gm cm$^2$) \\[15pt]
\hline

\hline
& 0.00 &2.355& 12.472&13.928 &1080.82 &1.502&1.986&2.259 &12.040&13.914 &1073.90&1.504&1.977\\[5pt]
      & 0.01 &2.350 &12.428& 13.854&1044.99 & 1.503&1.981&2.255&11.998 &13.836&1051.64 &1.505&1.966\\[5pt]
      NLD & 0.02&2.319 &12.154 &13.384&926.318 & 1.505&1.902&2.227&11.720 &13.356&907.157 &1.509&1.892\\[5pt]
      & 0.03 &2.243 &11.588&12.454 &669.213  & 1.515&1.748&2.161 &11.134&12.396&641.634&1.522&1.724\\[5pt]
      & 0.04 &2.118&10.712&11.306 &404.984 & 1.534&1.537&2.052&10.282 &11.206&371.858&1.545&1.500\\[5pt]
      & 0.05 &1.955 &9.718&10.154&217.511 &  1.564&1.313&1.910&9.540 &10.026&192.538 &1.579&1.272\\[5pt]
\hline
& 0.00 &1.922 &11.922&12.870 &682.506 & 1.529& 1.756&1.764 &10.084&11.210 &202.419&1.596&1.283\\[5pt]
      & 0.01 &1.917 &11.898&12.810 &663.885 & 1.530&1.745&1.762&9.990 &11.158&200.205 &1.597&1.279\\[5pt]
      IOPB & 0.02 &1.888&11.624 &12.442 &605.061 & 1.531&1.696 &1.744&9.860 &10.766&176.824 &1.599 &1.245\\[5pt]
      & 0.03 &1.815&10.978 &11.706 &458.665  & 1.538 &1.584&1.705&9.460 &10.254&148.144 &1.611&1.195\\[5pt]
      & 0.04 &1.697 &10.038&10.738 &291.840 & 1.559&1.406&1.634&8.918&9.502&102.871&1.641&1.102\\[5pt]
      & 0.05 &1.551 &8.992&9.612 &139.904& 1.616&1.180&1.535&8.244 &8.682&59.637 &1.712&0.982\\[5pt]
\hline

& 0.00 &1.764 &11.010&12.146 &457.878 &1.560& 1.579&1.607&9.164 &10.474&131.179 &1.696&1.157\\[5pt]
      & 0.01 &1.761 &10.972&12.090 &452.790 &1.561& 1.571&1.605&9.150 &10.428&129.365 &1.696&1.153\\[5pt]
      G3 & 0.02&1.738&10.704 &11.746 &406.223 & 1.562&1.528&1.593&9.006 &10.148&115.400 &1.699&1.123\\[5pt]
      & 0.03 &1.681 &10.156&11.070 &308.299  & 1.572&1.426&1.563&8.658 &9.584&86.460 &1.713&1.057\\[5pt]
      & 0.04 &1.589 &9.344&10.162 &187.794 & 1.602&1.263&1.509&8.182 &8.838&53.312 &1.749&0.956\\[5pt]
      & 0.05 &1.473 &8.406&9.014 &78.553 & 1.690&1.037&1.435&7.624 &7.958&25.068 &1.834&0.829\\[5pt]
\hline
\end{tabular}
\label{tab1}}
\end{sidewaystable}

\subsubsection{Correlation between moment of inertia of canonical star and dark matter Fermi momenta ($I_{1.4}=$$f(k_f^{DM})$):}
The moment of inertia can be calculated using the formula $I=MR^2$. In Fig. \ref{i14n}, we plot the moment of inertia of a canonical star \(I_{1.4}\) with dark matter Fermi momentum. We observe that the moment of inertia associated with dark matter decreases as the value of \(K_f^{DM} \) increases, as mentioned in Table \ref{tab1}. The presence of dark matter in the core of a neutron star softens its equation of state (EOS), resulting in a decrease in \( I_{1.4} \). Furthermore, we fit the \( I_{1.4}-K_f^{DM} \) curves and find a third-order polynomial correlation between the \( I_{1.4} \) and the parameter \( K_f^{DM} \). This correlation is expressed as:

\begin{equation}
    I_{1.4}/10^{45} = a (k_f^{DM})^3 + b (k_f^{DM})^2 + c (k_f^{DM}) + d.
\end{equation}

The coefficients \( a \), \( b \), \( c \), and \( d \) in this correlation vary depending on different parameter sets and compositions but the correlation function remain same. Thus, the relationship between \( I_{1.4} \) and \( K_f^{DM} \) is both parameter and composition-dependent.
\begin{figure}[t]
\begin{center}
\includegraphics[width=1\textwidth]{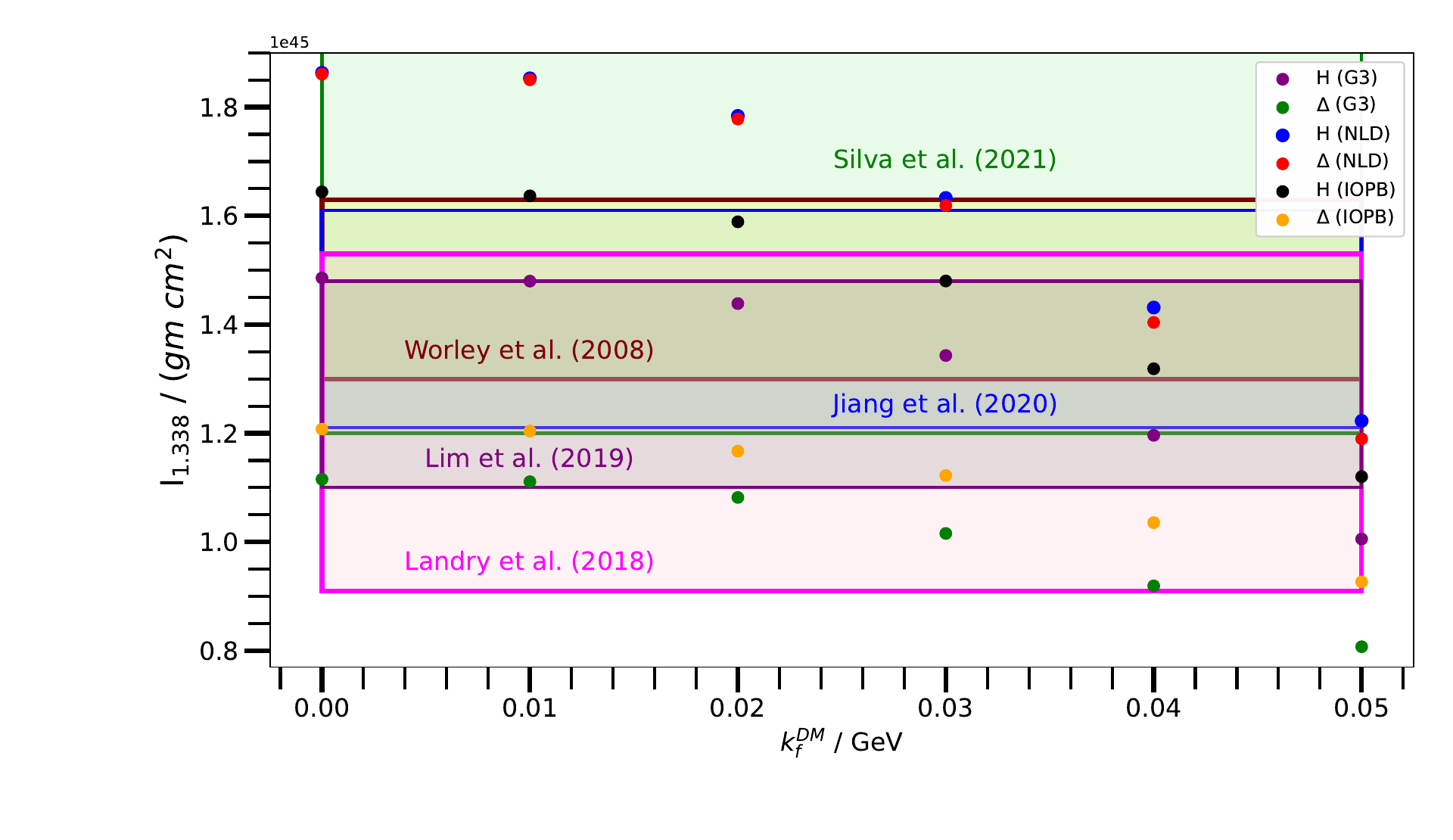}
\end{center}
\caption{ Moment of inertia of hyperon star corresponds to mass 1.338 $M_\odot$ ($I_{1.338}$) with NLD, G3, and IOPB parameter set for different values of dark matter Fermi momentum ($K_f^DM$= 0-0.05 GeV). The H represent the hyperon star without $\Delta-$baryons composition and $\Delta$ represent hyperon star with $\Delta-$baryons composition.}
\label{i1338}
\end{figure}

 Fig. \ref{i1338} shows the calculated values of moment of inertia ($I$) of PSR J0737-3039A pulsar with mass 1.338 $M_\odot$ and constraints of Worley et al. (2008) \cite{Worley_2008}: $I_{1.338}=$1.30--1.63 x 10$^{45}$ gm cm$^2$, Landry et al. (20018) \cite{Landry_2018}: $I_{1.338}=1.15^{+0.38}_{-0.24}$ x 10$^{45}$ gm cm$^2$, Lim et al. (2019) \cite{lim2019}: $I_{1.338}=1.36^{+0.12}_{-0.26}$ x 10$^{45}$ gm cm$^2$, Jiang et al. (2020) \cite{Jiang_2020}: $I_{1.338}=1.35^{+0.26}_{-0.14}$ x 10$^{45}$ gm cm$^2$ and Silva et al. (2021) \cite{silva2021}: $I_{1.338}=1.68^{+0.53}_{-0.48}$ x 10$^{45}$ gm cm$^2$. Comparing our result with these constraints, we find the G3 parameter set with hyperon composition satisfies all the mentioned constraints for $k_f^{DM}=0.00-0.03$ GeV. The NLD parameter set with $k_f^{DM}=0.04$ GeV satisfies all the constraints. The IOPB parameter set with hyperon composition also satisfies all the constraints for $k_f^{DM}=0.04$.

\section{Conclusion}
The universe is made up of 95\% dark matter and dark energy. Due to the strong gravitational pull of the neutron star, some of the dark matter components accreted inside the core of the neutron star. The dark matter content in the neutron star core affects the neutron star bulk properties, which depend on the nature of the dark matter candidate and how it interacts with neutron star matter inside the core. Here, we take the neutralino dark matter candidate which is a WIMP particle. We take the NLD parameter set of the RMF model for the calculation of nuclear bulk properties. The presence of dark matter softens the EOS; consequently, the maximum mass, corresponding radius, canonical radius, canonical tidal deformability, and moment of inertia decrease with $k_f^{DM}$.  However, the canonical frequency increases with $k_f^{DM}$.

In addition to that, we find a third-order polynomial correlation between dark matter Fermi momenta and neutron star bulk properties namely canonical radius, canonical tidal deformability, canonical frequency and moment of inertia of the canonical star. To check the parameter dependency of these correlations,  we take other two parameter set namely, G3 and IOPB and repeat the calculation. We find the functional form for all these correlations are same for each parameter set, however the correlation coefficients are vary for different parameter set. Furthermore, we alter the composition of the NS core by taking the hyperon matter with and without $\Delta$ composition. The functional form still same for the different composition of star but the correlation coefficients change. Our calculation suggests a correlation between dark matter fermi momenta and bulk properties of neutron star ( with or without $\Delta-$baryons) having a functional form of third-order polynomial. Further we try to constraint the DM content inside the NS core from the constraints of various properties such as $f_{1.4}$, $ \Lambda_{1.4}$ and $I_{1.338}$ given in various literature. Comparing our results with the various neutron star's constraints, we find that for $K_f^{DM} = 0.04$ and 0.05 GeV, our results align with most of the constraints. Detailed similar calculation with other properties may be investigated in the future.

\section{acknowledgments}
All calculations were supported by the Science and Engineering Research Board,
New Delhi, under project no. CRG/2022/005378.

\bibliography{dark}

\end{document}